# Quantized conductance and large *g*-factor anisotropy in InSb quantum point contacts

*Fanming Qu[†], Jasper van Veen[†], Folkert K. de Vries[†], Arjan J. A. Beukman[†], Michael Wimmer[†], Wei Yi[‡], Andrey A. Kiselev[‡], Binh-Minh Nguyen[‡], Marko Sokolich[‡], Michael J. Manfra[§], Fabrizio Nichele[♂], Charles M. Marcus[♂] and Leo P. Kouwenhoven[†,*]*

[†] QuTech and Kavli Institute of Nanoscience, Delft University of Technology, 2600 GA Delft, The Netherlands

[‡] HRL Laboratories, 3011 Malibu Canyon Rd, Malibu, CA 90265, USA

[§] Department of Physics and Astronomy, and Station Q Purdue

School of Electrical and Computer Engineering

School of Materials Engineering

Purdue University, West Lafayette, IN 47907, USA

[♂] Center for Quantum Devices, Niels Bohr Institute, University of Copenhagen, Copenhagen, Denmark

Keywords: InSb quantum well, quantum point contact, *g*-factor anisotropy, electron effective mass, conductance quantization

Due to a strong spin-orbit interaction and a large Landé *g*-factor, InSb plays an important role in research on Majorana fermions. To further explore novel properties of Majorana fermions, hybrid devices based on quantum wells are conceived as an alternative approach to nanowires. In this work, we report a pronounced conductance quantization of quantum point contact devices in InSb/InAlSb quantum wells. Using a rotating magnetic field, we observe a large in-plane ($|g_1| = 26$) and out-of-plane ($|g_1| = 52$) *g*-factor anisotropy. Additionally, we investigate crossings of subbands with opposite spins and extract the electron effective mass from magnetic depopulation of one-dimensional subbands.

Among the binary III-V semiconductors, InSb has the smallest effective mass and the highest room temperature mobility[1]. It further exhibits a strong spin-orbit interaction (SOI) and the largest Landé *g*-factor ($|g| = 51$ for the bulk), due to the strong coupling between the conduction band and the valence band resulting from the small energy gap[1-3]. Besides the continuously increasing interest in its various applications in spintronics[4], InSb has been extensively investigated for Majorana fermions and topological quantum computing (TQC)[5, 6]. Applying a magnetic field perpendicular to the spin-orbit field of a nanowire opens a Zeeman energy gap and creates one-dimensional (1D) helical states[7]. Furthermore, when a superconducting gap is induced through the proximity effect, a 1D topological superconductor can form and Majorana zero modes (MZMs) emerge at the boundaries of this topological phase. InSb nanowires played an important role in the first experimental signature of MZMs[8]. While rapid progress has been achieved based on InSb and InAs nanowires[9-12], further investigation of the non-Abelian properties and the unique fusion rules of MZMs requires more complex device designs[13-15]. Although crossed nanowires have been developed[16], simultaneously applying magnetic fields parallel to different branches of such nano-cross is difficult to realize, since this requires for example an "H" bar with parallel arms. Moreover, the scalability of nanowire systems for TQC can be challenging. Therefore, a "top-down" approach by fabricating 1D or network structures starting from a two-dimensional (2D) quantum well system is a promising alternative route.

InSb quantum wells have several important advantages over InSb nanowires. The mobility can exceed 200,000 cm$^2$/Vs[17-19], corresponding to a mean free path larger than 1.4 µm. To realize 1D helical states, it is crucial to optimize the potential profile for transport detection[20], which can be tailored through geometry design based on electrostatic modeling. In addition, the 2D electron gas (2DEG) functions as ideal contacts which naturally solves the interface problem for nanowires[21]. Although the calculated and reported Rashba SOI parameter α ranging 0.03 – 0.15 eVÅ[22-27] is smaller than that reported in nanowires[3, 28], a 2D heterostructure enables tuning of SOI strength[22, 23, 25-27, 29] by engineering asymmetric doping, barrier modulation and also electrical gating. The confinement of 1D structures defined on an InSb 2DEG may enhance α towards the values for nanowires. Together with the flexibility of complex device designs, these advantages motivate a detailed investigation of 1D structures based on InSb quantum wells. However, the successful gate depletion of InSb 2DEG was achieved only recently after solving the gate leakage problem[19, 30]. While 1D ballistic transport has been established in InSb nanowires and nanosails lately[31, 32], only one work reported an observation of quantized conductance in quantum point contacts (QPCs) on an InSb 2DEG[33]. Detailed transport properties including *g*-factors and the electron effective mass in 1D structures confined on InSb quantum wells are yet to be established. In this work, we demonstrate ballistic transport through QPCs in an InSb 2DEG. In a rotating magnetic field, the Zeeman spin splitting is investigated and a large in-plane and out-of-plane *g*-factor anisotropy is observed. Furthermore, crossings of subbands with opposite spins are studied and the electron effective mass is deduced using magnetic depopulation[34, 35].

The InSb/InAlSb heterostructure used in this work is grown on a GaAs (100) substrate using a fully relaxed $In_{1-x}Al_xSb$ buffer (x = 0.08). The quantum well consists of a 30 nm InSb layer sandwiched between $In_{1-x}Al_xSb$ barriers. Single side Si δ-doping sits 20 nm above the InSb layer in the top barrier. Details of the material growth and a full gate depletion of the 2DEG in Hall bar devices with $HfO_2$ as dielectric have been reported earlier[19]. To fabricate the QPC device studied here, as shown in Fig. 1a, a narrow constriction (~ 280 nm wide) on a 20 μm wide mesa is wet etched ~ 100 nm deep, followed by sputtering a 100 nm thick $Si_3N_4$ dielectric layer and evaporating a 100 nm wide Ti/Au top gate. The InSb 2DEG at both sides of the constriction functions as two in-plane side gates (SG). The Ohmic contacts, located > 20 μm away from the QPC, are formed by etching into the InSb layer using an Argon ion etch and in-situ deposition of Ti/Au layers. In addition, we also fabricated fully gate-defined QPCs, in which, instead of etching, the constriction is defined by two split Ti/Au side gates on top of $Si_3N_4$ dielectric and a global Ti/Au top gate on a second $Si_3N_4$ layer (see Supporting Information Fig. S1).

Transport measurements are carried out on the two types of InSb QPCs in both a cryo-free dilution refrigerator with a 6-2-1 T vector magnet and a $^3$He system with a single axis magnet of 9 T. Standard low frequency lock-in techniques are employed in a configuration of two-terminal or four-terminal measurement. Series resistances from the wires, measurement equipment and adjacent InSb 2DEG have been subtracted to match the quantized conductance for the data reported below, unless otherwise stated. For the source-drain bias spectroscopy, the voltage drop on the QPC is also corrected accordingly. Hall bar devices with $Si_3N_4$ as dielectric are characterized before performing the QPC measurements (see Supporting Information Fig. S2). By a comparison of the two types of QPCs, we find that the etch-defined QPC shows pronounced quantized conductance plateaus at zero magnetic field, while the fully gate-defined type requires a small perpendicular magnetic field to suppress backscattering and interference. Therefore, we focus on the former in the following and briefly present the results on the latter in the Supporting Information Fig. S1.

Differential conductance $G = dI/dV = I_{ac}/V_{ac}$ is measured by applying a small ac excitation voltage $V_{ac}$ with or without a dc bias voltage $V_b$ and measuring the ac current $I_{ac}$. Figure 1b shows $G$ as a function of side gate voltage $V_{SG}$ with a fixed top gate voltage $V_{TG} = 0.3$ V at different temperatures. Quantized conductance plateaus at $nG_0$ are observed resulting from the ballistic transport in the 1D constriction, where $n$ = 1, 2 and 3, and $G_0 = 2e^2/h$ ($h$ is the Planck constant and $e$ the elementary charge). At mixing chamber temperature $T = 26$ mK, small conductance fluctuations indicate finite backscattering and interference processes around the QPC. The rest of the data reported below are all measured at 26 mK. As shown in Fig. 1c, $G$ can be controlled by both top gate and side gates, confirming their proper functioning. Figure 1d displays the numerically calculated derivative of $G$ with respect to side gate voltage, *i.e.*, the transconductance $dG/dV_{SG}$, versus $V_{SG}$ and $V_b$. (Raw data of $G$ versus $V_{SG}$ and $V_b$ is shown in Fig. S3b of the Supporting Information.) As indicated by the green solid arrows, subband spacings $E_{2-1}$ and $E_{3-2}$ of ~4.6 meV are roughly equal ($E_{i-j}$ represents the energy spacing between the $i^{th}$ and $j^{th}$ subbands where $i$ and $j$ are an integer), suggesting to a near-parabolic confinement potential.

We further examine the spin splitting of the 1D subbands in a magnetic field. Figure 2a shows $G$ as a function of $V_{SG}$ and $B_x$ (along current flow) where $n$ can now assume half integer values. As $B_x$ increases from 0 T, half integer plateaus resulting from Zeeman spin splitting appear and widen in $V_{SG}$ while the integer plateaus narrow down. The evolution of spin resolved subbands in $B_x$ is illustrated in the right panel of Fig. 2b ignoring SOI. At $B_x = 3$ T, as shown by the green curve in the left panel of Fig. 2b, only half integer plateaus survive when the spin-down band from the $i^{th}$ subband (i↓) crosses the spin-up band from the $(i+1)^{th}$ subband ((i+1)↑). When $B_x > 3$ T, after the crossing of the spin split subbands, the spin-up bands 1↑ and 2↑ are the lowest two bands in energy. At $B_x = 4$ T, the integer plateaus are restored but are now fully spin polarized for $1G_0$. Note that a combination of a large $g$-factor and the modest subband separation enables such clear crossing at a moderate magnetic field[36, 37]. When two 1D subbands of opposite spins cross, a spontaneous spin splitting and the emergence of the so called 0.7 analog at the 1.5 plateau have been reported in GaAs 2DEG[36], which, as well as the $0.7G_0$ feature, are absent here in our InSb QPCs but require further investigation.

At large $B_z > 1$ T (out-of-plane), as shown in Fig. 3a, in contrast to the case of $B_x$, all plateaus widen due to Zeeman splitting and magnetic depopulation of 1D subbands, as will be discussed below. For the case of $B_y$ (in-plane but perpendicular to current flow), as displayed in Fig. 3b, the behavior is similar to that in $B_x$, although here the measured magnetic field range is smaller.

To directly inspect the evolution of the spin splitting in a magnetic field along different orientations, the magnetic field is rotated in the x-z plane (Fig. 3c) and the x-y plane (Fig. 3d) while keeping the amplitude fixed at 1.8 T and 1 T, respectively. The magnetoresistance from the adjacent InSb 2DEG increases as the $B_z$ component rises. And thus, after subtracting a constant series resistance at $B_z = 0$, the calculated conductance at finite $B_z$ is lower than the actual value. Consequently, in the regions labelled by 0.5 and 1 in Fig. 3c, the quantized plateaus drop below $0.5G_0$ and $1G_0$, respectively (see Supporting Information Fig. S5a). A noteworthy feature when the $B_z$ component increases is that both the 0.5 and 1 plateaus widen in $V_{SG}$. Assuming a constant gate voltage to energy conversion, the Zeeman splitting in the first subband, $E_{1\downarrow-1\uparrow} = |g_1|\mu_B B$ with $\mu_B$ the Bohr magnetron, is proportional to the width of the 0.5 plateau along the gate voltage axis. Figure 3c thus shows a g-factor anisotropy up to a factor of ~2 between the z and x directions (see Supporting Information Fig. S5b). In contrast, the in-plane (x-y plane) g-factor is nearly isotropic as suggested by the roughly constant width of the 0.5 plateau in Fig. 3d (see Supporting Information Fig. S6a).

To determine the magnitude of the g-factor quantitatively, source-drain bias spectroscopy is performed. Figures 4a, 4b and 4c show the numerically calculated transconductance $dG/dV_{SG}$ as a function of $V_{SG}$ and $V_b$ at $B_z = 2$ T, $B_z = 1.5$ T and $B_x = 2$ T, respectively, with a fixed $V_{TG} = 0.5$ V. The black dashed lines are guides to the eye and help to read out the energy spacings as marked by the green solid arrows. From $E_{1\downarrow-1\uparrow} = |g_1|\mu_B B$, the effective g-factor for the first subband can be extracted to be $|g_{1,z}| \approx 51$ ($B_z = 2$ T), $|g_{1,z}| \approx 53$ ($B_z = 1.5$ T) and $|g_{1,x}| \approx 26$ ($B_x = 2$ T), exhibiting an anisotropy, as already indicated by Fig. 3c. We would like to emphasize that the difference in absolute values between $|g_{1,z}|$ and $|g_{1,x}|$ of ~26 is large. Such knowledge of g-factor anisotropy is important for future experiments on helical states and MZMs. One direct consequence is that the Zeeman energy changes differently for different magnetic field orientations. Consistent with Fig. 3d, the extracted $|g_{1,y}| \approx 28$ is close to $|g_{1,x}|$, showing a nearly isotropic in-plane *g*-factor (see Supporting Information Fig. S6). In 2D quantum wells, the effective electron *g*-factor becomes anisotropic due to lower symmetry introduced by the heterostack. It is also renormalized (usually reduced) owing to subband confinement and strain[38-41]. The 1D constriction may further modify the effective *g*-factors, though, the extracted anisotropy is larger than theoretical calculations[38, 39]. The effective *g*-factor for the second subband can also be obtained from $E_{2\downarrow\text{-}2\uparrow}$ giving $|g_{2,z}| \approx 38$ ($B_z = 1.5$ T) and $|g_{2,x}| \approx 23$ ($B_x = 2$ T), both smaller than the first subband, in qualitative agreement with experimental results on InSb nanowire quantum dots[3], albeit the obtained magnitude of renormalization is somewhat unexpected here. Detailed theoretical discussions are supplied in the Supporting Information (Section III).

Since the extracted *g*-factor anisotropy of $|g_{1,z}|/|g_{1,x}| \approx 2$ from the bias spectroscopy agrees well with the anisotropy suggested by the width of the 0.5 plateau ($\Delta V_{SG}$) in Fig. 3c, the assumption of the constant gate voltage to energy conversion is supported. Therefore, we can use $\Delta V_{SG}$ to deduce the *g*-factor at different angles in the x-z plane. Figure 5a shows the transconductance $dG/dV_{SG}$ of Fig. 3c, where the white-red color represents the transition between conductance plateaus. The green arrow illustrates the width of the 0.5 plateau $\Delta V_{SG}$. The *g*-factor for the first subband $|g_1|$ can be now obtained from $\left|g_{1,z}\right| \times \Delta V_{SG}(angle) / \Delta V_{SG}(angle = 90°)$ with $|g_{1,z}| \approx 52$. Figure 5b presents the angular anisotropy of $|g_1|$.

Next we turn to magnetic depopulation to extract the electron effective mass. At $B = 0$, the parabolic confinement from the gates (as indicated by the fact that $E_{2\text{-}1} \approx E_{3\text{-}2}$ in Fig. 1e) results in subband spacings of $\hbar\omega_0$ ($\hbar = h/2\pi$). When a perpendicular magnetic field (along $\vec{z}$) is applied, an additional magnetic parabolic potential enhances the level separation to $\hbar\sqrt{\omega_0^2 + \omega_c^2}$, where $\omega_c = eB_z/m^*$ is the cyclotron frequency ($m^*$ is the effective mass)[34, 35, 42]. Hence, at finite $B_z$, $m^*$ can be extracted from the subband spacing $E_{2\text{-}1}(B_z) = 1/2E_{1\downarrow\text{-}1\uparrow} + E_{2\uparrow\text{-}1\downarrow} + 1/2E_{2\downarrow\text{-}2\uparrow} = \hbar\sqrt{\omega_0^2 + \omega_c^2}$. Neglecting the orbital effect of $B_x$, $\hbar\omega_0 \approx 4.7$ meV ($\omega_c = 0$) using the energy intervals in Fig. 4c at $B_x = 2$ T ($E_{1\downarrow\text{-}1\uparrow} \approx 3.0$ meV, $E_{2\uparrow\text{-}1\downarrow} \approx 1.9$ meV, $E_{2\downarrow\text{-}2\uparrow} \approx 2.6$ meV). Consequently, the effective mass is calculated to be $m^* \approx 0.017m_e$ and $0.019m_e$ by applying the energy separations at $B_z = 1.5$ T and 2 T, respectively, with $m_e$ the electron mass. More details of the calculation can be found in the Supporting Information (Section II). For an InSb quantum well, confinement enhances the effective mass and the nonparabolicity of the band dispersion enhances it further at finite densities[39]. The average $m^*$ of $0.018m_e$ is larger than the bulk value of $0.014m_e$[1], but is consistent with theoretical calculations for a 30 nm thick InSb quantum well at low densities[39].

In conclusion, we demonstrate a high quality conductance quantization of QPCs on InSb quantum wells. In a rotating magnetic field, Zeeman spin splitting is investigated and a large in-plane and out-of-plane g-factor anisotropy is observed. In a moderate in-plane magnetic field, clear crossings of electron subbands with opposite spins are achieved. Moreover, for the first time, the electron effective mass is extracted from magnetic depopulation of 1D subbands in InSb QPCs. Further research on InSb quantum wells on carefully designed hybrid devices is needed to pursue helical states and Majorana zero modes.

AUTHOR INFORMATION

**Corresponding Author**

* E-mail: L.P.Kouwenhoven@tudelft.nl

**Notes**

The authors declare no competing financial interest.

ACKNOWLEDGMENT

We gratefully acknowledge Michal Nowak, Jakob Kammhuber and Maja Cassidy for very helpful discussions. This work has been supported by funding from the Netherlands Foundation for Fundamental Research on Matter (FOM) and Microsoft Corporation Station Q.

ASSOCIATED CONTENT

**Supporting Information**

Extra figures, analysis and theoretical discussions. This material is available free of charge via the Internet at http://pubs.acs.org.

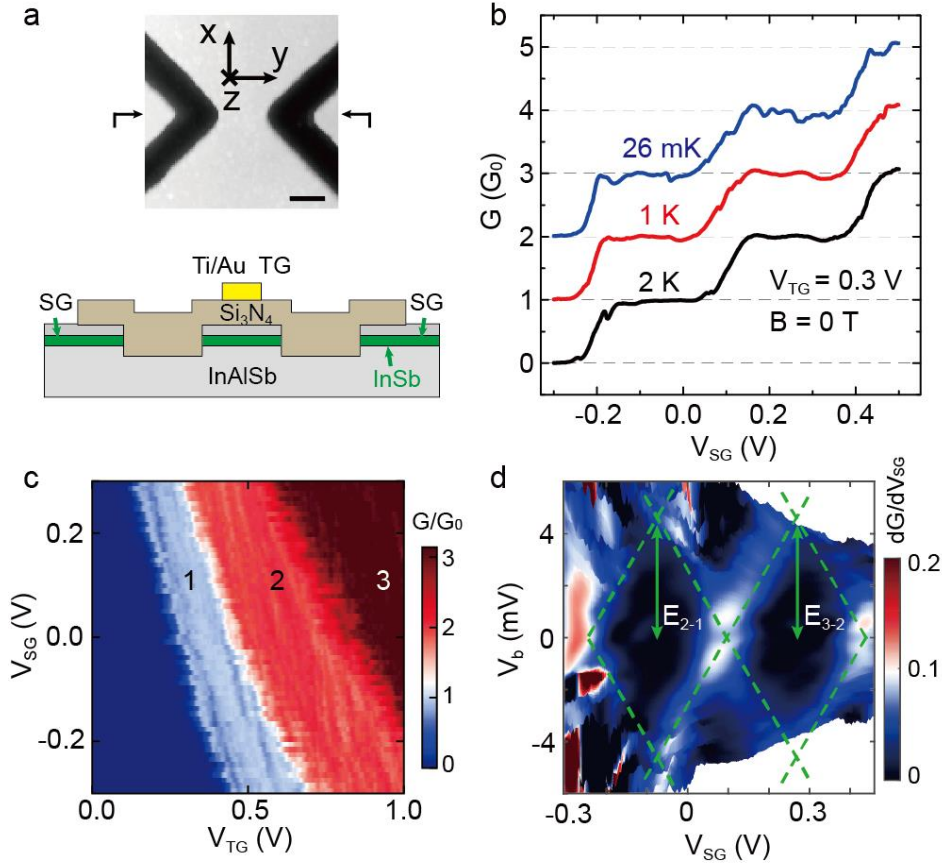

**Figure 1:** Conductance quantization in etch-defined InSb QPCs. (**a**) Image and schematics of the etch-defined QPC device. The top panel shows an atomic force microscope image of the constriction after a wet etch (before depositing $Si_3N_4$ and Ti/Au layers) with the black region being ~100 nm deep. The scale bar is 200 nm. The axes illustrate the vector magnet orientations. The bottom panel displays the cross-section of the device along the plane marked by the two arrows in the top panel. The constriction is controlled by two etch-defined in-plane side gates (SG) and a 100 nm wide top gate (TG). (**b**) Differential conductance $G$ versus side gate voltage $V_{SG}$ curves at a fixed top gate voltage $V_{TG} = 0.3$ V for different temperatures. Traces are offset by $1G_0$ ($G_0 = 2e^2/h$) for clarity. (**c**) $G$ as a function of both $V_{TG}$ and $V_{SG}$. (**d**) Numerically calculated transconductance $dG/dV_{SG}$ as a function of $V_{SG}$ and dc bias voltage $V_b$ at $V_{TG} = 0.3$ V. The green dashed lines are guides to the eye and the green solid arrows indicate the subband spacings.

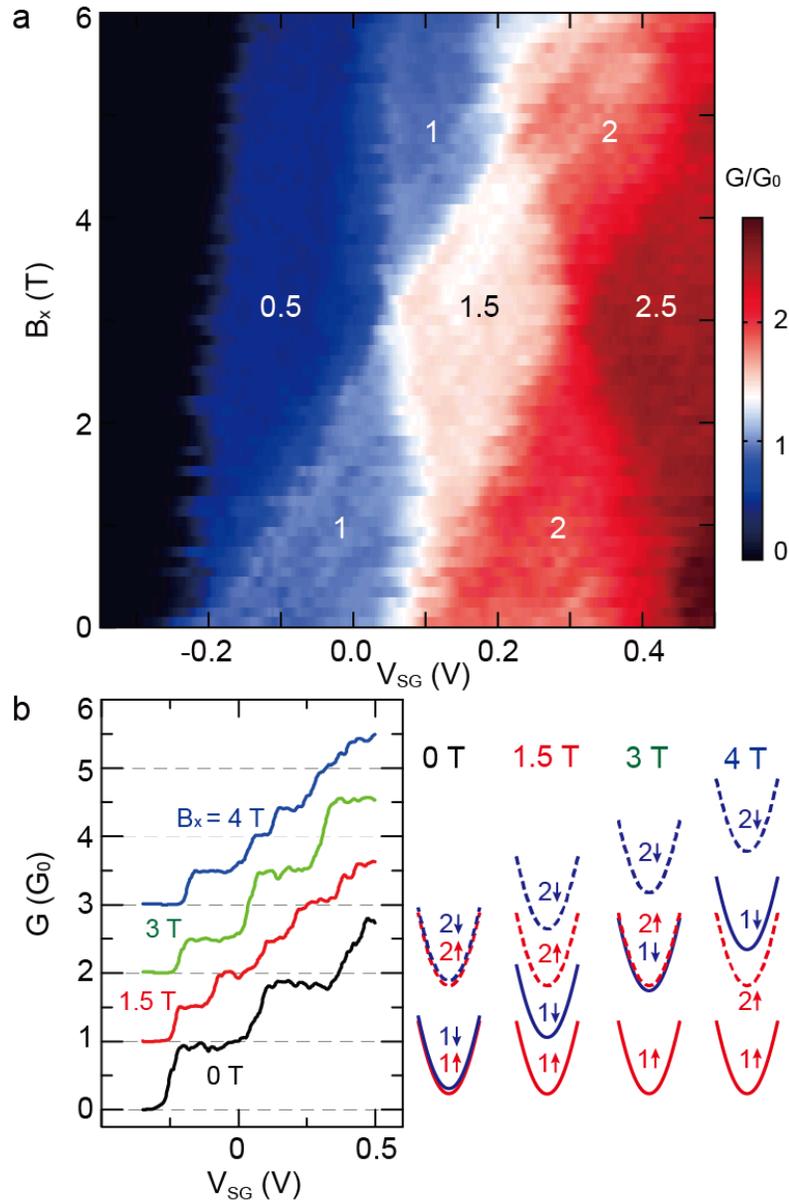

**Figure 2:** Crossings of electron subbands with opposite spins. (**a**) $G$ versus $V_{SG}$ and $B_x$ (along current flow) at $V_{TG} = 0.3$ V with numbers $n$ labelling quantized conductance at $nG_0$. (**b**) The left panel shows line cuts taken from (**a**) at different magnetic fields. Traces are offset by $1G_0$ for clarity. The right panel displays band dispersions at different $B_x$, sketching the evolution of the spin resolved subbands.

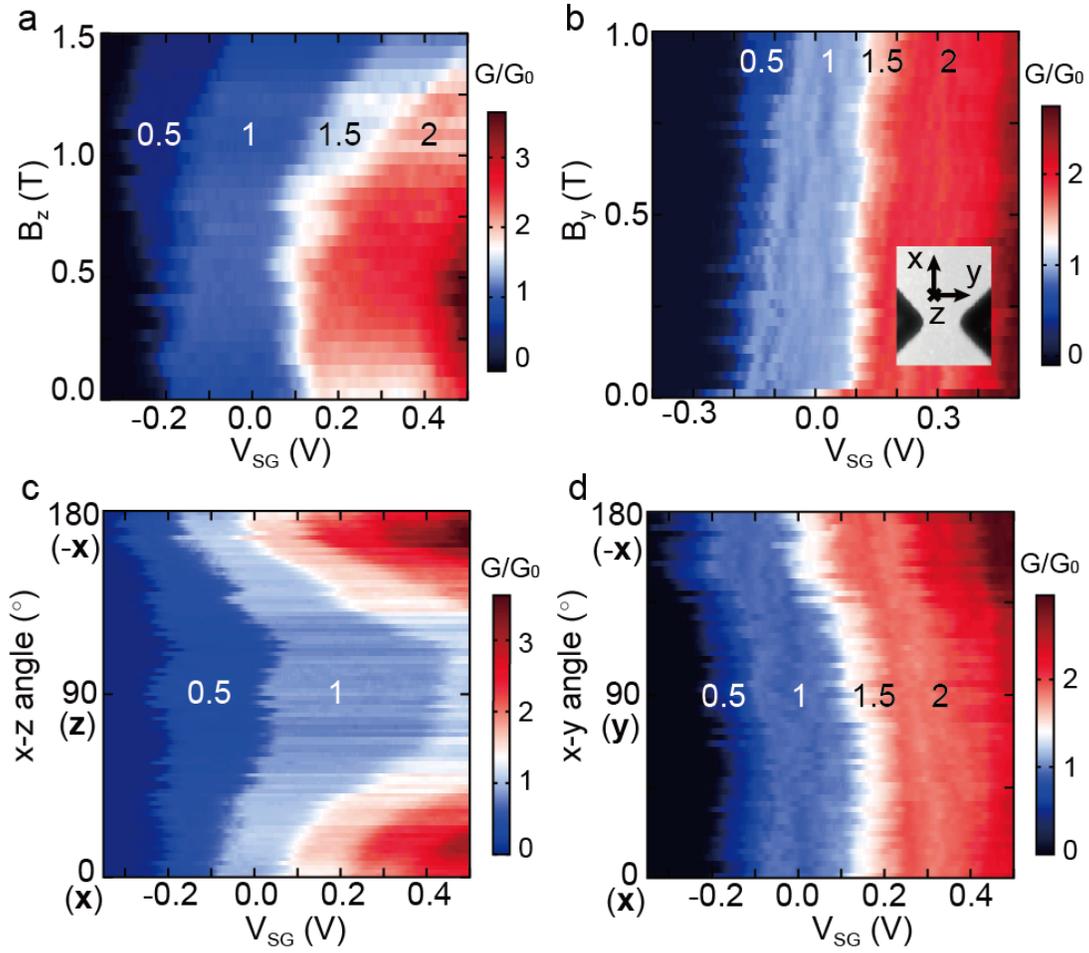

**Figure 3:** Spin splitting in magnetic fields. (**a**) $G$ versus $V_{SG}$ and $B_z$, (**b**) $G$ versus $V_{SG}$ and $B_y$ with numbers $n$ marking quantized conductance at $nG_0$ ($V_{TG}$ = 0.3 V). (**c**, **d**) $G$ as a function of $V_{SG}$ and the x-z angle at a fixed magnetic field amplitude of 1.8 T (**c**), and the x-y angle at a fixed magnetic field amplitude of 1 T (**d**). The x-z angle = 0, 90 and 180 degrees correspond to magnetic field along $\vec{x}$, $\vec{z}$ and $-\vec{x}$, respectively. Accordingly, in the x-y plane these three angles stand for $\vec{x}$, $\vec{y}$ and $-\vec{x}$ directions.

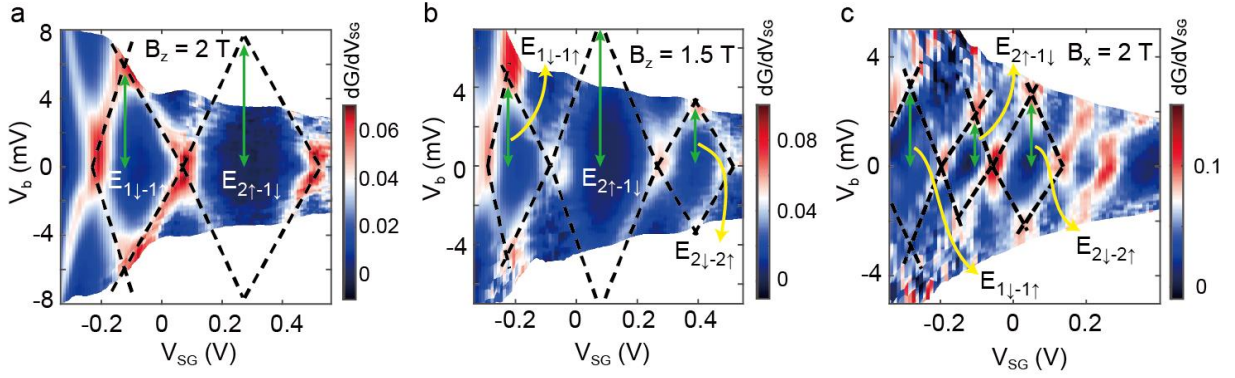

**Figure 4:** Bias spectroscopy and *g*-factors. Transconductance d*G*/d*V*_SG as a function of *V*_SG and *V*_b at (**a**) $B_z$ = 2 T, (**b**) $B_z$ = 1.5 T, and (**c**) $B_x$ = 2 T with a fixed $V_{TG}$ = 0.5 V. The larger d*G*/d*V*_SG (red color) represents transitions between quantized conductance plateaus. Black dashed lines are guides to the eye and green solid arrows indicate the level spacings.

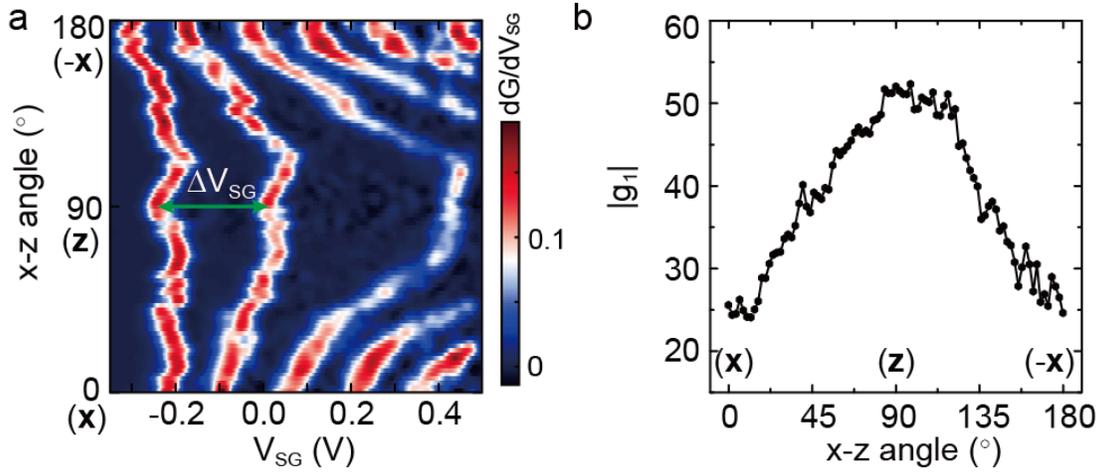

**Figure 5:** Anisotropic *g*-factors. (**a**) Transconductance d*G*/d*V*_SG extracted as the numerical derivative of the data in Fig. 3c. The green arrow indicates the 0.5 plateau width $\Delta V_{SG}$. (**b**) First subband *g*-factor $|g_1|$ in the x-z plane calculated based on $\Delta V_{SG}$ (see text).

# Supporting Information for "Quantized conductance and large *g*-factor anisotropy in InSb quantum point contacts"

*Fanming Qu[†], Jasper van Veen[†], Folkert K. de Vries[†], Arjan J. A. Beukman[†], Michael Wimmer[†], Wei Yi[‡], Andrey A. Kiselev[‡], Binh-Minh Nguyen[‡], Marko Sokolich[‡], Michael J. Manfra[§], Fabrizio Nichele[♂], Charles M. Marcus[♂] and Leo P. Kouwenhoven[†,*]*

[†] QuTech and Kavli Institute of Nanoscience, Delft University of Technology, 2600 GA Delft, The Netherlands

[‡] HRL Laboratories, 3011 Malibu Canyon Rd, Malibu, CA 90265, USA

[§] Department of Physics and Astronomy, and Station Q Purdue
School of Electrical and Computer Engineering
School of Materials Engineering
Purdue University, West Lafayette, IN 47907, USA

[♂] Center for Quantum Devices, Niels Bohr Institute, University of Copenhagen, Copenhagen, Denmark

*Email: l.p.kouwenhoven@tudelft.nl

**Section I: Extra figures**

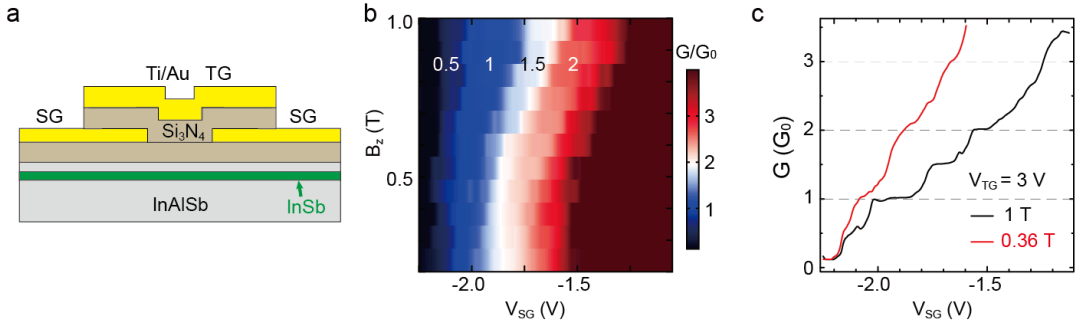

**Figure S1**: Results on fully gate-defined QPCs. (a) Cross-section of the fully gate-defined InSb QPC. The constriction is defined by two split side gates separated by 400 nm on top of the $Si_3N_4$ dielectric (100 nm thick) and a global top gate. Figure S1b shows a typical color plot of $G$ (the unit $G_0 = 2e^2/h$) versus $V_{SG}$ and $B_z$ at a fixed $V_{TG} = 3$ V (270 nm of $Si_3N_4$ in total) and $T = 0.3$ K. In general, this type of QPC requires a small perpendicular magnetic field to suppress conductance fluctuations associated with backscattering and interference, presumably owing to the large effective 2DEG area and interface area to define the QPC. Figure S1c illustrates two line cuts taken from Fig. S1b (series resistance has been subtracted for each trace) showing quantized conductance in $B_z$. By a comparison of the quality of quantized conductance plateaus, the etch-defined QPC shown in the main text is preferred for further research on helical states and Majorana zero mode related physics.

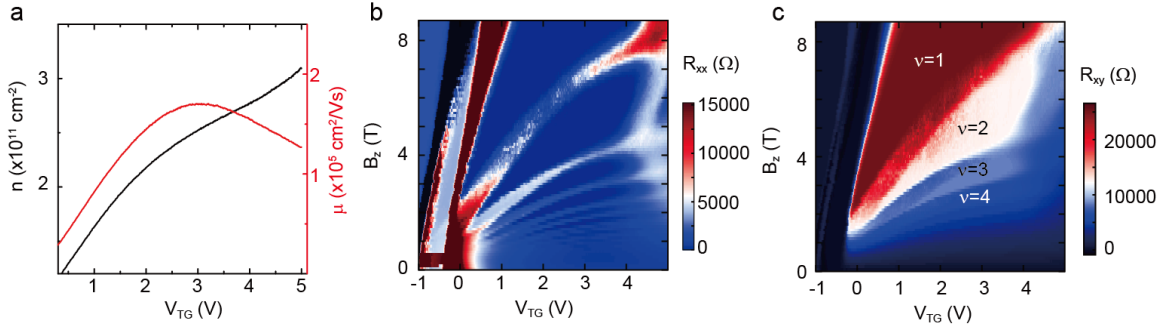

**Figure S2**: Characterization of an InSb Hall bar device. To fabricate the Hall bar device, 20 μm wide mesa is first wet etched, followed by sputtering a 150 nm thick $Si_3N_4$ dielectric layer and a Ti/Au top gate. The Ohmic contacts are formed by etching into the InSb layer using an Ar ion etch and an in-situ deposition of Ti/Au layers. Transport measurements are performed in a $^3$He system at 300 mK. (a) Density and mobility as a function of top gate voltage $V_{TG}$. The mobility drops around $V_{TG}$ = 3 V due to occupancy of the second subband. More details can be found in a previous work[1]. (b) Longitudinal resistance $R_{xx}$ and (c) Hall resistance $R_{xy}$ as a function of both $V_{TG}$ and out-of-plane magnetic field $B_z$. At high magnetic field, integer quantum Hall effect is observed where $R_{xx}$ reaches zero resistance (dark blue region in (b)) and quantized Hall resistance forms, as marked by the filling factors in (c). These results are similar as that with $HfO_2$ as dielectric on the same wafer[1].

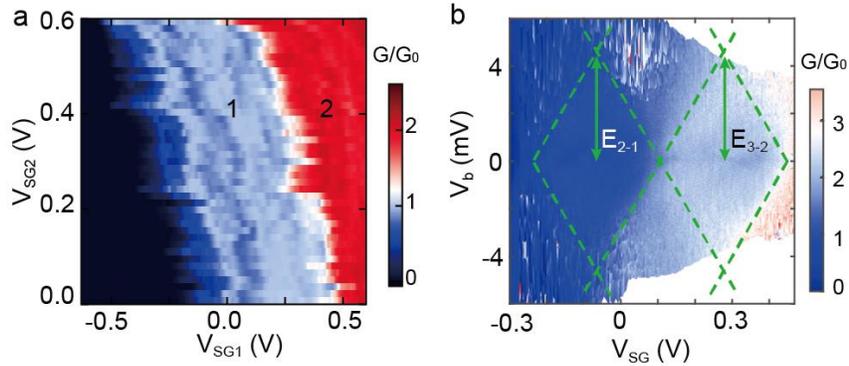

**Figure S3**: Differential conductance $G$ at $B$ = 0 for the etch-defined QPC shown in the main text. (a) $G$ as a function of gate voltage $V_{SG1}$ and $V_{SG2}$ on each of the two in-plane side gates, at a fixed $V_{TG}$ = 0.5 V. (b) $G$ versus source-drain bias voltage $V_b$ and $V_{SG}$ ($V_{SG} \equiv V_{SG1} = V_{SG2}$) at a fixed $V_{TG}$ = 0.3 V. The numbers $n$ = 1, 2 in (a) represent quantized conductance at $nG_0$. Note that the pinch-off voltage in (a) (the boundary between black and blue) presents a slope of ~ -1 ($\Delta V_{SG2}/\Delta V_{SG1} \approx$ 0.6 V/(-0.5 V)), suggesting nearly symmetric side gates. The green dashed lines are guides to the eye, marking the transitions between quantized conductance plateaus. Transconductance shown in Fig. 1d in the main text is numerically calculated from the data in (b). Level separations are labelled by $E_{2\text{-}1}$ and $E_{3\text{-}2}$.

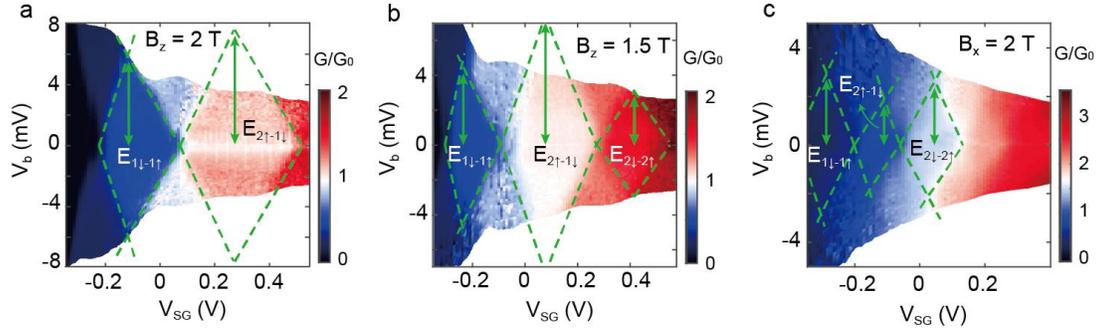

**Figure S4**: Differential conductance $G$ for the etch-defined QPC shown in the main text. $G$ as a function of $V_{SG}$ and $V_b$ at (a) $B_z = 2$ T, (b) $B_z = 1.5$ T, and (c) $B_x = 2$ T at a fixed $V_{TG} = 0.5$ V. The numerically calculated transconductance shown in Fig. 4 in the main text is calculated based on these plots accordingly. Green dashed lines are guides to the eye and green solid arrows illustrate level spacings, the same as that in Fig. 4 in the main text.

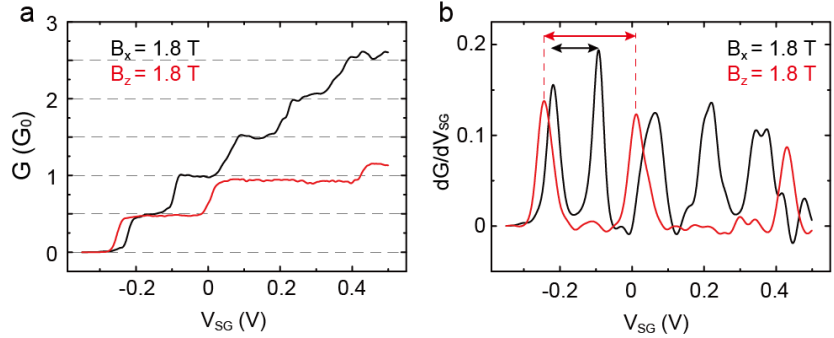

**Figure S5**: Line cuts taken from Fig. 3c in the main text. (a) Two line cuts at x-z angle = 0 (along current flow, $B_x = 1.8$ T, black) and 90 ($B_z = 1.8$ T, red) degrees. For the whole 2D color plot of Fig. 3c a series resistance of 6.5 kΩ from the adjacent InSb 2DEG is subtracted to match the quantized conductance at $B_x = 1.8$ T (black). However, when the $B_z$ component increases, the magnetoresistance of the adjacent 2DEG rises and thus the calculated conductance of the QPC drops lower than the actual value. As shown by the red line, the conductance at the quantized plateaus is below the expected value. Nevertheless, the width of the 0.5 plateau at $B_z = 1.8$ T is wider than that at $B_x = 1.8$ T, indicating a $g$-factor anisotropy. The exact width can be obtained from the numerically calculated transconductance $dG/dV_{SG}$ as shown by the red and black arrows in (b), $\Delta V_{SG} = 0.256$ V and 0.126 V, respectively. Therefore, an out-of-plane and in-plane $g$-factor anisotropy of ~2 can be extracted.

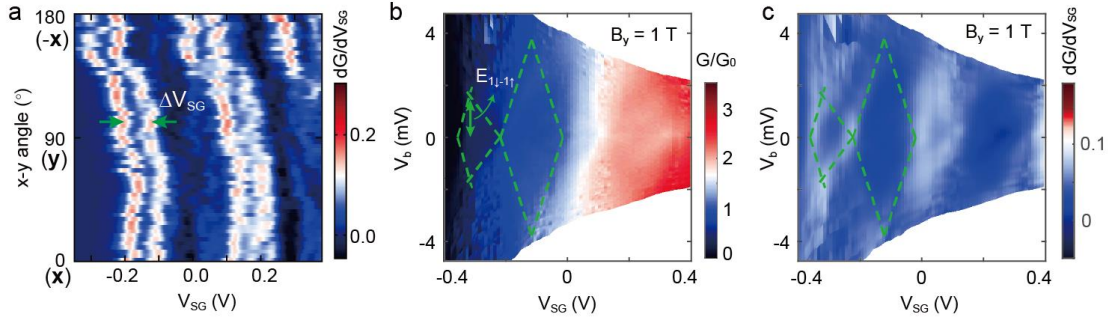

**Figure S6**: g-factors in $B_y$ (in-plane but perpendicular to current flow) for the etch-defined QPC in the main text. Figure 4 in the main text shows the bias spectroscopy of the QPC transconductance in $B_x$ and $B_z$, from which anisotropic g-factors are extracted. As shown in Fig. 3d in the main text, the roughly constant width of the 0.5 plateau when the magnetic field is rotated in the z-y plane implies an isotropic in-plane g-factor. This figure represents (a) the numerical transconductance of Fig. 3d in the main text, (b) bias spectroscopy in $B_y$ = 1 T and (c) its numerical transconductance. The high $dG/dV_{SG}$ in (a) (red color) represents transitions between conductance plateaus and the green arrow illustrates the width of the 0.5 plateau $\Delta V_{SG}$, which is roughly constant. Note that due to the low field along this direction of the vector magnet, as guided by the green dashed lines and solid arrows in (b) and (c), the energy separation $E_{1\downarrow\text{-}1\uparrow}$ is small (~ 1.6 meV) leading to a relatively large uncertainty. But still, from $E_{1\downarrow\text{-}1\uparrow} = |g_{1,y}|\mu_B B$, $|g_{1,y}| \approx 28$, close to $|g_{1,x}| \approx 26$ as shown in the main text.

## Section II: Calculation of the electron effective mass

In this section we present the details of the calculation of the electron effective mass. As discussed in the main text, at $B=0$, the parabolic confinement from the gates results in subband spacings of $\hbar w_0$ ($\hbar = h/2\pi$). When a perpendicular magnetic field (along $\vec{z}$) is applied, an additional magnetic parabolic potential enhances the level separation to be $\hbar\sqrt{\omega_0^2 + \omega_c^2}$, where $\omega_c = eB_z/m^*$ is the cyclotron frequency ($m^*$ is the effective mass). At a fixed $B_z$,

$$E_{1\text{-}2}(B_z) = 1/2 E_{1\downarrow\text{-}1\uparrow} + E_{2\uparrow\text{-}1\downarrow} + 1/2 E_{2\downarrow\text{-}2\uparrow} = \hbar\sqrt{\omega_0^2 + \omega_c^2} \ . \quad \text{(Eq. 1)}$$

When $B_z=0$, neglecting the orbital effect of $B_x$, $\hbar\omega_0 = E_{1\text{-}2}(B_z = 0, B_x = 2\text{ T}) \approx 4.7$ meV using the energy intervals in Fig. 4c in the main text at $B_x=2$ T ($E_{1\downarrow\text{-}1\uparrow} \approx 3.0$ meV, $E_{2\uparrow\text{-}1\downarrow} \approx 1.9$ meV, $E_{2\downarrow\text{-}2\uparrow} \approx 2.6$ meV), consistent with $E_{1\text{-}2}(B=0) \approx 4.6$ meV deduced in Fig. 1d in the main text at $V_{TG} = 0.3$ V.

At $B_z = 1.5$ T, by plugging the subband spacings from Fig. 4c ($E_{1\downarrow\text{-}1\uparrow} \approx 4.6$ meV, $E_{2\uparrow\text{-}1\downarrow} \approx 7.3$ meV, $E_{2\downarrow\text{-}2\uparrow} \approx 3.3$ meV) into Eq. 1, the effective mass is calculated to be $m^* \approx 0.017 m_e$ with $m_e$ the electron mass.

When $B_z = 2$ T, as shown in Fig. 1a in the main text, the 1.5 plateau is outside the applied $V_{SG}$ range because of the wide plateaus. However, $E_{2\downarrow\text{-}2\uparrow}$ can be calculated using $|g_{2,z}|$ obtained at $B_z=1.5$ T and the corresponding $m^* \approx 0.019 m_e$ at $B_z = 2$ T.

**Section III: Theoretical discussions**

Here, we discuss results of model calculations of the electron g-factor tensor at the bottom of the first electron subband formed in the InSb quantum well (QW) layer in our device. With device mesa naturally cut along [110], [-110] axes, the electron g-factor tensor is expected to be diagonal in the chosen (x, y, z) coordinate system, so as the Zeeman part of the electron spin Hamiltonian can be written as

$$H_Z = \frac{\mu_B}{2} \sum_{\alpha=x,y,z} \sigma_\alpha g_{\alpha\alpha} B_\alpha$$.

In a planar QW, $g_{xx} \approx g_{yy} \neq g_{zz}$ define the in-plane component and the component along growth axis, respectively. The full theoretical procedure and resulting explicit formulas for g-factor components are provided in Ref. 2. Following this procedure, as a first step, we find the quantized electron state in the framework of the 8x8 **k·p** model that accounts exactly for the coupling between the lowest conduction band $\Gamma_6$ and the upper valence bands $\Gamma_8$ and $\Gamma_7$ and also retains all remote band terms that notably affect the electron dispersion. Then, we evaluate the in-plane g-factor component for the calculated electron state using Eq. (6) of Ref. 2 and the g-factor along the heterostack growth axis with the help of Eq. (10) of Ref. 2. In doing so, we specifically account for the exact InSb/InAlSb heterostack, strain, and biasing in presence of the fixed and self-consistent mobile charges (for details, see Ref. 1). Material parameters used in present calculations are taken, without exception, from Ref. 3. Most parameters characterizing the band structure of the InSb are well established (notable exception is the partitioning of the strain-induced shifts between conduction and valence bands, and especially details of the valence band deformation potentials). For InAlSb alloys, we interpolate between InSb and AlSb using recommended bowing constants, when available, and default to linear interpolation otherwise.

We obtain, in the limit of small B and zero in-plane wave vector, a value of -37.3 (-36.2) for the in-plane component of the electron g-factor and a value of -46.3 (-46.0) for the component along growth axis when biasing device to the onset of the QW conduction, most relevant in QPC experiments (at the onset of the second conducting channel — see Ref. 1 for details) — both components *up* from the value of -51 in bulk InSb. In addition to observing the minimal effect of the device biasing on g-factor components, we have verified that sensitivity to plausible uncertainties in heterostack specifications and material parameters is also modest (with valence band deformation potentials being the biggest suspects as mentioned above).

For the QPC intersubband spacing of 4.6 meV we estimate in-plane span of ~60 nm for the QPC ground subband — larger, but actually still comparable to the QW width of ~30 nm. Thus, electron states in the QPC are wire-like. Effects of the wire-like confinement on electron *g*-factor were analyzed in detail in Ref. 4. In addition to slightly renormalizing all *g*-factor components further *up*, i.e., further away from the bulk value, in-plane confinement should also lead to *weak* in-plane anisotropy $g_{xx} \neq g_{yy}$ (see, e.g., Fig. 3 in Ref. 4).

Obviously, correspondence between calculation and experiment is somewhat wanting — unlike calculation, *g*-factor measured along growth axis is at (or even exceeding) the InSb bulk value, and *g*-factor anisotropy, while in a qualitative agreement, is notably smaller in calculation. The issue requires further analysis, meanwhile we would like to mention two intricacies: (i) Anticipated *g*-factor renormalization by many-body effects, including those due to 2DEG finite spin polarization (issue somewhat addressed in Ref. 5), While it is expected to lead primarily to larger absolute *g*-factor values (even exceeding originating bulk numbers), it is probably not going to explain large anisotropy observed experimentally. (ii) Contributions of Rashba (R) and Dresselhaus (D) spin-orbit interactions to the electron spin Hamiltonian

$$H_s = H_Z + H_R + H_D$$

when considering states with a finite wave vector — e.g., at the Fermi level, when quantized by finite *B*, and/or resulting from QPC in-plane confinement. With our choice of (x, y, z) coordinates in respect to the crystallographic directions,

$$H_R = \alpha(\sigma_x k_y - \sigma_y k_x) \text{ and } H_D = \beta(\sigma_x k_y + \sigma_y k_x)$$

in a planar QW (in a linear-in-k order). These additional terms could affect observed spin splittings thus distorting extracted *g*-factor values.

For our device, we have evaluated a Rashba coefficient α ~ 14 meVÅ at the onset of the QW conduction; α is up to 36 meVÅ at the onset of the second channel due to stronger biasing and Dresselhaus coefficient β of up to 78 meVÅ. Using, as a relevant example, $k_F$ corresponding to the full QPC intersubband spacing of 4.6 meV, resulting Rashba and Dresselhaus spin splittings $2\alpha k_F \leq 0.25$ meV and $2\beta k_F \leq 0.54$ meV, respectively. They are to be compared to Zeeman splitting $|g|\mu_B B = B \times 2.9$ meV/T (when employing, roughly, bulk InSb |g| = 51).

For *B* along the growth axis, the Zeeman and SO spin splittings sum quadratically, and, at $B > 1$ T, the Zeeman terms are already too large and easily overwhelm the SO terms. Thus the *perceived* value of *g*-factor along the growth axis is not going to be affected by the SO terms. For the in-plane *B*, they can be directly additive, and experimentally extracted *g*-factor values could be influenced by the SO terms even for $B > 1$ T.